\documentclass[seceq]{ptptex}

\usepackage{graphicx}
\usepackage{epsfig}

\newcommand{\be}{\begin{equation}}
\newcommand{\ee}{\end{equation}}
\newcommand{\bea}{\begin{eqnarray}}
\newcommand{\eea}{\end{eqnarray}}
\newcommand{\ba}{\begin{array}}
\newcommand{\ea}{\end{array}}

\newcommand{\Tr}{\mbox{Tr}}

\newcommand{\p}{\partial}

\newcommand{\ep}{\epsilon}

\newcommand{\cL}{{\cal{L}}}
\newcommand{\cM}{{\cal{M}}}

\newcommand{\cS}{{\cal{S}}}

\newcommand{\nn}{\nonumber}
\newcommand{\dder}[1]{\frac{\partial}{\partial #1}}
\newcommand{\drdr}[2]{\frac{\partial #1}{\partial #2}}

\title{
Improved Renormalization Group Analysis 
for Yang-Mills Theory%
}

\author{
Shoichi \textsc{Kawamoto}$^1$ %
and
Toshihiro \textsc{Matsuo}$^2$
}

\inst{
$^1${\it
Rudolf Peierls Centre for Theoretical Physics, University of Oxford, \\
1 Keble Road, Oxford, OX1 3NP, UK
}\\
$^2${\it
Department of Physics,
National Taiwan Normal University, \\
Taipei City, 116, Taiwan
}\\
}

\recdate{September 26, 2007}

\abst{
We apply an improved renormalization group analysis for pure
Yang-Mills theory in four dimensions in the Landau gauge and $O(N)$ nonlinear sigma model
in two dimensions.
We find a pole in the gluon two-point function at one-loop order,
which is generated non-perturbatively.
In Yang-Mills theory this may correspond to the expectation value of a
gauge invariant operator and emerges as
$M_\text{P}/\Lambda_{\overline{MS}} \simeq 0.81$,
where $\Lambda_{\overline{MS}}$ is the $\overline{MS}$ scale.

}

\begin{document}

\maketitle

\section{Introduction}

The improved perturbation theory formulated by Gell-Mann and Low
\cite{Gell-Mann:fq} is a theoretical framework with which, using the ideas
of the renormalization group with the results of perturbation theory
to a given order, one can determine something about the next order of
perturbation theory.
Employing the same type of philosophy, variational approximation methods,
by which one can extract certain nonperturbative information from the
results of perturbation theory, have been developed in both theoretical
and numerical approaches.

The essential idea of the variational approximation scheme, which is
often called an improved mean field approximation in the literature,
 can be
summarised simply by the expression ``the principle of minimal
sensitivity''. \cite{Stevenson:1981vj}
Suppose that we have a Lagrangian $\cL$ of interest.
For concreteness, we use the $\phi^4$ theory as an example.
We first introduce a mean-field Lagrangian $\cL_{\rm m}=m^2\phi^2/2$,
where $m^2$ is a parameter to be tuned such that the quantity
we would like to evaluate is independent of this parameter.
We then rewrite the original Lagrangian as
\begin{align}
    \cL=&
    \frac{1}{2}\left( \partial_\mu \phi \right)^2
    -\frac{m_0^2}{2}\phi^2 - \frac{g}{4!}\phi^4 \nn\\
     & \Rightarrow \
    \frac{1}{2}\left( \partial_\mu \phi \right)^2
    -\frac{m^2}{2}\phi^2 + \lambda\left(
     -\frac{m_0^2}{2}\phi^2+\frac{m^2}{2}\phi^2 -\frac{g}{4!}\phi^4 
\right)\,,
\end{align}
where we have introduced the formal coupling constant $\lambda$.
We then regard the first two terms as the unperturbed action and the
terms in the parentheses as a perturbation with respect to $\lambda$,
and calculate the quantity of interest using perturbation theory.
This $\lambda$ will be set to $1$ at the end of calculation, and thus,
in principle, any physical quantity calculated with this Lagrangian
cannot depend on $m^2$,
but, as we will see, the $m^2$ dependence remains if terms of higher
order in $\lambda$ are dropped.
Finally, we tune the parameter $m^2$ for the calculated quantity to be independent
of it as should be the case.
In other words, one should choose the parameter for the quantity of
interest to be on a plateau whose emergence signals that the
approximation scheme works well.
This is the meaning of the expression ``the minimal sensitivity''.

When we apply the variational approximation
 for a massless theory, it
is convenient to start with the massive counterpart
$\mathcal{L}_{\text{massive}}(m^2,g)$ to the massless theory and
employ the ordinary perturbation theory with respect to the coupling
$g$.
All we have to do is to replace the mass and coupling as $m^2
\rightarrow (1-\lambda)m^2$ and $g \rightarrow \lambda g$ in the
result of the perturbation theory and then keep the desired order in
the formal coupling constant $\lambda$ and set $\lambda=1$. 
Here, the mass plays the role of the parameter to be tuned.


There have been many works carried out with this method.
For the case of anharmonic oscillators, it was shown that
the new perturbation series obtained with this recipe is convergent\cite{Delta_conv}.
(See also Ref. \citen{DeltaExp}.)
Some progress has been made in the study of matrix models and
reduced Yang-Mills models
\cite{Kabat:2000hp,Sugino:2001,Nishimura:2001sx,KKKMS1,Nishimura:2003gz,Nishimura:2004ts,Aoyama:2005nd,Aoyama:2006di,Aoyama:2006rk,Aoyama:2006je,Aoyama:2006yx}.
In Ref. \citen{Nishimura:2002va}, $D$-dimensional pure Yang-Mills
theory is studied in the context of the reduced model and
convergence of Monte Carlo data is demonstrated.
In Ref. \citen{Nishimura:2001sx} and, subsequently, Ref. \citen{KKKMS1},
the IIB matrix model, which is defined as the reduced maximally supersymmetric Yang-Mills
theory and is also conjectured to be a nonperturbative
definition of superstring theory, is studied, and the emergence of
four-dimensional space-time as a vacuum of string theory is
suggested.\footnote{
In Ref. \citen{Aoyama:2006di}, higher-order calculations are given
 and the result strengthens the original idea.
(See also Ref. \citen{Aoyama:2006je} and references therein.)}

Regarding applications to quantum field theory, in Ref. \citen{AGN-AGIKN}
the variational method is combined with the renormalization group
method.
The mass gap of the Gross-Neveu model in two dimensions is calculated
and is found to exhibit good agreement with the exact solution.
Furthermore, in a subsequent series of papers \cite{AGKN-K},
the method is applied to QCD in order to study, among other things, the dynamical origin
of the quark mass.
In this paper, we call the method developed there ``the improved
renormalization group analysis'', and  we review it in $\S$
\ref{sec:VM_method} with some refinement.

This improved mean field approximation can be
regarded as an ``improved Taylor expansion'', which is a general
scheme to improve the convergence of a Taylor expansion series of
interest\cite{KKKMS1}.
We note here that when we attempt to apply the improved mean field
approximation to quantum field theories, we usually face the problem
of determining at which stage we should make the replacement,
i.e., how to make it compatible with the renormalization procedure.
This problem arises because both involve the choice of the unperturbed
part of the action, and usually they do not seem to be compatible.
The proposal made in  \citen{AGN-AGIKN} and \citen{AGKN-K} is that
the replacement is carried out for the bare Lagrangian, as the
original mean field approximation suggests, and then the same
renormalization scheme as for the original Lagrangian is applied.
It was shown that this leads a consistent improved series;
that is, the divergences are removed by this renormalization.
Here we employ another approach.
Inspired by the spirit of the improved Taylor expansion,
we first apply usual renormalization scheme and improve the quantity of
interest by using a renormalization group technique, and then apply the
improved Taylor expansion to functions that admit a Taylor expansion.
Surprisingly, our method gives the same improved
function as the method in Refs. \citen{AGN-AGIKN} and \citen{AGKN-K}.
We therefore believe that our general scheme still can be interpreted as
an improved mean field approximation.

In this paper, we apply the improved renormalization group analysis to
the pure Yang-Mills theory in four dimensions in order to explore
the nonperturbative nature of the theory, such as confinement.
Though lattice theory provides a method for nonperturbative study, there is still
a great need for analytic methods that we could trust.
Recently, there has been growing interest in the condensate of a
mass dimension two operator in QCD and Yang-Mills theory.
Such condensation is believed to be a key to understanding the
confinement problem.
\cite{GSZ,Kondo:2001pb,Boucaud:2001st,A2condensates}${}^{,}${}\footnote{For
  recent developments, see, e.g., Refs. \citen{Browne:2006uy}, \citen{Dudal:2006xd}
 and references therein.}
The operator considered here is  
$\Delta=\frac{1}{2}(\text{volume})^{-1}\langle \text{min}_U \int d^4 x
(A_\mu^U)^2 \rangle$, where $U$ represents a gauge transformation, (volume)
denotes the space-time volume, and ``min'' represents the operation of
 taking minimum with respect to gauge transformations.
This operator is indeed gauge invariant, but non-local in a general gauge.
In the Landau gauge, this is the expectation value of the
gauge-variant 
local operator $\int \left\langle \frac{1}{2}A_\mu A^\mu
\right\rangle$ \cite{A2condensates}.
Then, in the Landau gauge, the mass term of the gluons, $\Tr \ m^2 A_\mu A^\mu$,
might be induced nonperturbatively through the four-point interaction.{}
\footnote{%
Note that no clear explanation has been given of the connection between
mass generation in gauge theory and the condensate of the
operators.
There have been several arguments regarding this issue.
(See Refs. \citen{Kondo:2001pb} and \citen{Barnich:ve}.)
Here we simply assume the mass generation of the Landau gauge.}
This ``induced mass'' may appear as a pole of the gluon propagator.
Thus, we perform the improved renormalization group analysis
to evaluate the mass pole of the gluon propagator in the Landau gauge.
We do not demand that this ``mass'' be the physical mass of the gluon, 
since the mass term is not gauge invariant.
However, what we calculate is assumed to be the
expectation value of a gauge invariant quantity.
Therefore, though this does not have the meaning of a mass, except in
the Landau gauge, the quantity itself is thought to have a gauge invariant
meaning.
In other words, we simply propose a possible way to calculate $\Delta$ in Yang-Mills theory.
The condensation of these kinds of operators has been discussed by means
of lattice simulations \cite{Boucaud:2001st} as well as from
a phenomenological point of view \cite{GSZ,A2condensates}. 

This paper is organized as follows.
In the next section we explain the improved renormalization group
method developed in Refs. \citen{AGN-AGIKN} and \citen{AGKN-K}, with our own
refinement, employing the idea of the improved Taylor expansion.
Then, in $\S$ \ref{sec:NLS} we demonstrate an application to the $O(N)$
nonlinear sigma model in two dimensions. 
In $\S$ \ref{sec:YM} we calculate the mass pole of the gluon
propagator in the Landau gauge by using the Curci-Ferrari model
\cite{Curci:bt} as a massive counterpart of Yang-Mills theory.
We perform the calculation in a general covariant gauge, in which we can
obtain the result in the Landau gauge by setting the gauge parameter $\xi$
to zero.
Section \ref{sec:conc} is devoted to conclusions and discussion.

\section{Improved renormalization group analysis}
\label{IRGM}

We first describe an improved renormalization group analysis
that we subsequently apply to Yang-Mills theory on the basis of
Refs. \citen{AGN-AGIKN} and \citen{AGKN-K}.
We start with a Lagrangian density $\mathcal{L}(m^2,g,\mu)$, where
$m^2$ is the mass parameter, $g$ is a dimensionless coupling constant,
and $\mu$ is a mass scale which is introduced to keep the coupling $g$
dimensionless.
This massive theory should be regarded as the massive counterpart of a
massless theory of interest.
As we have seen in the introduction, in order to obtain an improved
perturbation series in $\lambda$ for a given quantity, such as propagators,
in the massless theory,
it is sufficient to replace $m^2$
with $(1-\lambda)m^2$ and $g$ with $\lambda g$ in the quantity
calculated using perturbation theory with the massive action.
After setting $\lambda=1$, one can, at least naively, expect that this
new quantity describes that of the massless theory.
Therefore, we first review the method for obtaining a pole mass, in which we are
interested here, improved by a renormalization group equation in the
massive theory.
As stated in the introduction, we realize this substitution by use of
the improved Taylor expansion, which is in general
expected to improve the convergence of a series and to extract
information regarding the original function from its perturbation series of
\textit{finite} degree.
We can easily apply this method to a perturbation series in quantum
field theory, even after 
we have improved the perturbation series by use of the renormalization
group technique.
In this case, we simply improve the quantity of interest, which
is written as a series in a coupling constant, by replacing the
original parameters
with new ones, including a formal expansion parameter denoted by
$\lambda$ in the introduction.
Here we note that there can be some quantities that do not admit
a Taylor expansion, for example, a function of a fractional power in the coupling
constant appearing through a nonperturbative refinement like the
renormalization group.
We do \textit{not} consider such quantities and will improve only the
parts that can be written as a power series with integer powers in the
coupling constant.
This is the general criterion for the improvement we propose in this
paper.
This procedure provides the same formula as that proposed in
Refs.~\citen{AGN-AGIKN}  and \citen{AGKN-K}, where the improve
mean-field approximation is
applied to the bare Lagrangian, and then the renormalization and the
renormalization group method are applied carefully.
We thus conclude that our method is actually an improved approximation of
a massless theory of interest.

Here, let us start by explaining how to improve the pole mass expression
in which we are interested in this paper.
As explained in Appendix \ref{LeadingLog}, a pole mass can be written in
terms of these parameters as
\begin{equation}
         M^2_{\text P} = m^2 \sum_{n=0}^{\infty} \sum_{L=0}^{n}
         A_{n,L} \left[ \ln \left(\frac{m^2}{\mu^2} \right)
         \right]^{n-L} g^{2n} , 
\end{equation}
where the quantities $A_{n,L}$ are constant, that is,
are independent of $m^2$, $\mu$ and $g$.
The coefficients with $L<n$ are governed by the renormalization group
equation and can be determined recursively
with the condition $A_{0,0}=1$, while the coefficients
$A_{n,n} (\equiv A_n)$ need to be fixed by perturbative
calculations and are called ``non-logarithmic corrections.''

If we set the scale $\mu$ equal to $m$,
which is accomplished at the specific point
\bea
\cM^2=m^2(\mu=\cM) ,
\eea
then all the logarithmic terms vanish, and we obtain
\begin{equation}
       M_\text{P}^2 = m^2(\cM) \left( 1+ A_1 g^2(\cM) + A_2
g^4(\cM)+\cdots
\right),
\end{equation}
where we denote the running mass and coupling constant at the scale $\mu$
as $m^2(\mu)$ and $g^2(\mu)$, respectively.
Although both $g^2(\cM)$ and $m^2(\cM)$ should be determined by
the full renormalization group equation (RGE),
we can approximate them by replacing them with $g^2_\text{1-loop}$ and
$m^2_\text{1-loop}$, which are solutions to the RGE
at one-loop order, namely
\bea
m^2_\text{1-loop}=\frac{m^2(\mu)}
{\left[1+b_0g^2(\mu)
\ln(\frac{m_\text{1-loop}^2}{\mu^2})\right]^{\frac{\gamma_0}{2b_0}}} ,
\quad
g^2_\text{1-loop}=\frac{g^2(\mu)}
{1+b_0g^2(\mu)\ln(\frac{m_\text{1-loop}^2}{\mu^2})} ,
\eea
where $b_0$ and $\gamma_0$ are the coefficients of
the renormalization group functions $\beta(g^2)=-2b_0 g^4 - O(g^6)$ and
$\gamma_m(g^2)=\gamma_0 g^2 +O(g^4)$.
Thus we have
\begin{equation}
    \label{Mp_1loop}
         M^2_{\text P} = m_\text{1-loop}^2
         (1 + g_\text{1-loop}^2 A_{1} + g_\text{1-loop}^4 A_{2} +
         \cdots) .
\end{equation}
According to Appendix \ref{LeadingLog},
this approximation corresponds to the leading-logarithm
approximation with perturbative non-log corrections.
At this stage, we assert that the approximation is valid if the running
coupling constant $g^2_\text{1-loop}$ remains small at $\mu=\cM$.
Thus, we can apply this approximation to asymptotically free theories,
if $\cM$ is large enough.
Thus we should choose the initial value of the running mass  $m^2(\cM)$
in order for these conditions to be satisfied.

\subsection{One-loop improved RG analysis}
\label{sec:VM_method} 

Let us consider a pole mass at one-loop order in ordinary
perturbation theory,
\begin{equation}
    M_\text{P}^2=m^2(\bar{\mu})\left[
      1-\frac{\gamma_0}{2}g^2(\bar{\mu}) \ln
\left(\frac{m^2}{\bar{\mu}^2}\right)
      +A_1 g^2(\bar{\mu})+\mathcal{O}(g^4) \right],
\end{equation}
where $\gamma_0$ must be such that this expression is consistent with the expression
(\ref{Mp_1loop}), and $A_1$, which comes from a finite part in the
renormalization prescription, is the non-logarithmic correction.
Note that we have used a ``mass-independent renormalization scheme'',
like the $MS$ or $\overline{MS}$ scheme.

The leading logarithm contributions can be included as explained above,
and then the pole mass becomes
\bea
M_P^2&=&
m_\text{1-loop}^2(m)(1 + g_\text{1-loop}^2(m^2) A_{1})
\nn \\
&=&\frac{m^2(\bar{\mu})}{\left[1+b_0 g^2
\ln(\frac{m_\text{1-
loop}^2}{\bar{\mu}^2})\right]^{\frac{\gamma_0}{2b_0}}}
+m_\text{1-loop}^2(m)g_\text{1-loop}^2(m)A_{1} .
\eea
It is convenient to rewrite this as
\bea
M^2_P=
\frac{m^2(\bar{\mu})}{(b_0g^2(\bar{\mu})F)
^{\frac{\gamma_0}{2b_0}}}
\left[1+\frac{A_{1}}{b_0F}\right] ,
\eea
where
\bea
F=\frac{1}{b_0g^2(\bar{\mu})}+\ln
\left(\frac{m_\text{1-loop}^2}{\bar{\mu}^2}\right) .
\eea
Note that $F$ satisfies the recursive relation
\bea
F=-\ln\left([b_0g^2(\bar{\mu})F]^{\frac{\gamma_0}{2b_0}}
\frac{\Lambda_{\overline{MS}}^2}{m^2(\bar{\mu})}\right) \,,
\eea
with the basic scale
$\Lambda_{\overline{MS}}^2=\bar{\mu}^2
e^{-\frac{1}{b_0 g^2}}$,
where $\bar{\mu}^2=4\pi e^{-\gamma}\mu^2$.
Let us define the dimensionless parameter
$x=(b_0 g^2)^{-\frac{\gamma_0}{2b_0}}m^2
/\Lambda_{\overline{MS}}^2$.
Then the expression for the pole mass and
the above defining equation for $F$ become
\bea
M^2_P=\Lambda^2_{\overline{MS}}
\frac{x}{F^{\frac{\gamma_0}{2b_0}}}
\left[1+\frac{A_{1}}{b_0F}\right] ,
\eea
and
\bea
F=-\frac{\gamma_0}{2b_0}\ln F+\ln x .
\eea
Note that this function has a logarithmic cut starting from $x=0$ in
the complex $x$-plane.
We choose this cut so that $F$ is analytic for $x>0$.

Now we have a renormalization group expression for the pole mass
at one-loop order, and the next task is to apply the variational method
around a massless theory in the manner reviewed in the introduction.

\begin{figure}[t]
\begin{center}
\includegraphics[width=25em]{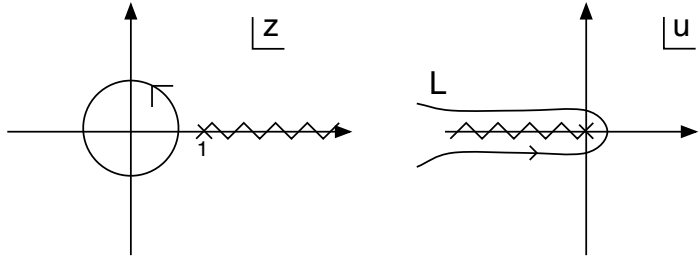}
\end{center}
\caption{Contour path.}
\label{contour path}
\end{figure}

The improved renormalization group analysis for the massless theory
is realized by the substitution
\bea
m^2 \to (1-\lambda)m^2 ,\quad g^2 \to \lambda g^2 ,
\eea
in the parts that admit power series expansions of integer powers in
$\lambda$.
Note that we have improved the pole mass expression by use of the
renormalization group, and actually its recursive form is a result of
this refinement. 
As a result, there are some terms in $M_P^2$ that do not admit
a Taylor expansion.
Because the coupling constant appears in a nonperturbative fashion,
it is not difficult to see that it is sufficient for the improvement to
replace $x$ with $(1-\lambda) x$ and to leave the rest unchanged.
Then the mass $M_P^2$ comes to depend on the formal coupling
$\lambda$ and is expanded in power series in $\lambda$ as
\bea
M_P^2(\lambda)=  \sum_{m=0}^\infty a_m \lambda^m .
\eea
We define the $n$-th order mass after setting $\lambda=1$ as
\bea
M^{2}_{(n)} &=& \sum_{m=0}^n a_m
\nonumber \\
&=&
\oint \frac{dz}{2\pi i}
\left( \frac{1}{z}+\cdots +\frac{1}{z^{n+1}}\right) M_P^2(z)
\nonumber \\
&=&
\Lambda^2_{\overline{MS}} x \oint \frac{dz}{2\pi i}
\left(-1+z^{-(n+1)}\right)
\frac{1}{F^{\frac{\gamma_0}{2b_0}}}
\left[1+\frac{A_{1}}{b_0F}\right]  .
\eea
Note that the analyticity of $F$ leads to the conclusion that
$[F((1-z)x)]^{-\frac{\gamma_0}{2b_0}}$ and
$[F((1-z)x)]^{-\frac{\gamma_0}{2b_0}-1}$ cannot be singular
at the origin of the complex $z$ plane when $x>0$,
and therefore the first term in the integrand must vanish.
Let us define $u \equiv n(1-z) $.
Then we have
\bea
M^{2}_{(n)} =
\Lambda^2_{\overline{MS}} x \int_L \frac{du}{2\pi i}
\left(1-\frac{u}{n}\right)^{-(n+1)}
\frac{1}{F^{\frac{\gamma_0}{2b_0}}}
\left[1+\frac{A_{1}}{b_0F}\right] .
\eea
Here, the contour path $L$ is taken around the cut on the
negative real axis in the $u$-plane, as shown in Fig.~\ref{contour path}.

The approximation becomes better as higher-order terms are taken in the
perturbative expansion in the formal coupling $\lambda$.
Thus it is natural to take the limit $n \to \infty$, and we then obtain
\bea
M^{2} \equiv \lim_{n \to \infty}M^{2}_{(n)} =
\Lambda^2_{\overline{MS}} x \int_L \frac{du}{2\pi i} e^u
\frac{1}{F^{\frac{\gamma_0}{2b_0}}}
\left[1+\frac{A_{1}}{b_0F}\right] .
\eea
   From the defining relation,
$F$ can be expanded around $x=0$ (hence $u=0$)
and we have
\bea
[F(u)]^{-B}
=(xu)^{-1}\left[1+(xu)^{1/B}
+\frac{B-2}{2B}(xu)^{2/B}
+O((xu)^{3/B}) \right] ,
\eea
where $B=\gamma_0/(2b_0)$.
We find an approximate expression for the pole mass
with the help of the formula
\bea
\frac{1}{\Gamma(z)}=\int_L \frac{dt}{2\pi i}e^t t^{-z} \,.
\eea
This expression is
\begin{align}
\frac{M^{2}(x)}{\Lambda_{\overline{MS}}^2} =&
1+\frac{1+B}{B}\frac{A_1}{b_0}
+\frac{A_1}{b_0}\frac{x^{-1/B}}{\Gamma\left(1+\frac{1}{B}\right)}
+\left(1+\frac{B^2-1}{2B^2}\frac{A_1}{b_0}
\right)\frac{x^{1/B}}{\Gamma\left(1-\frac{1}{B}\right)} \nn\\
&+\left( \frac{B-2}{2B}+\frac{B^3-3B^2+4}{6B^3}\frac{A_1}{b_0} \right)
\frac{x^{2/B}}{\Gamma\left(1-\frac{2}{B}\right)}
+\cdots .
\end{align}

We would like to find a plateau with respect to the parameter
$x$,\footnote{%
Recall that $x$ corresponds to the mass parameter to be tuned.}
because the dynamically induced
mass can be accurately approximated by the value
on such a  plateau.

We can further refine the above result by introducing a
scale-changing parameter $a$, which
is defined as a scaling parameter that rescales
$\bar{\mu}$ to $a \bar{\mu}$.
Although the rescaling of a renormalization point does not
affect the renormalization invariant quantities, such as the
one-loop mass $m^2_\text{1-loop}$,
the perturbative mass does depends on the parameter $a$.
However, the pole mass must be independent of $a$,
and thus it is natural to tune $a$ in such a way that a plateau clearly emerges.
Once we introduce the parameter $a$, the defining equation for $F$
becomes
\bea
F=\ln\left(\frac{xu}{a}\right)-\frac{\gamma_0}{2b_0}\ln F .
\eea
Finally, we arrive at
\bea
\label{eq:IRG_formula}
\frac{M^2_P}{\Lambda_{\overline{MS}}^2}=
x \int_L \frac{du}{2\pi i}
e^u F^{-\frac{\gamma_0}{2b_0}}
\left[1+\left(\frac{A_{1}}{b_0}+\frac{\gamma_0}{2b_0}\ln
a\right)\frac{1}{F}
\right] .
\label{polemassformula}
\eea
Thus, if we know $b_0$, $\gamma_0$ and $A_1$, we can calculate
$M_\text{P}^2 / \Lambda_{\overline{MS}}^2$ as a function of $x$ and
$a$.
We then search for a plateau of this function with respect to $x$ by
varying the parameter $a$ and take the value on the plateau as the
approximated value of $M_\text{P}^2 / \Lambda_{\overline{MS}}^2$.

\section{Improved RG analysis for the nonlinear sigma model}
\label{sec:NLS}
Having reviewed the basic strategy, we move on to the $O(N)$
nonlinear sigma model in two dimensions as a preliminary example.
The $O(N)$ invariant Lagrangian density with an external field is
\begin{equation}
  \mathcal{L}= \frac{1}{2g^2} \left(
\partial_\mu \boldsymbol{S}\cdot \partial^\mu \boldsymbol{S}
 - 2 \boldsymbol{h}\cdot \boldsymbol{S}
 \right) \,,
\end{equation}
where $\boldsymbol{S}$ is the sigma model field of magnitude $1$ (i.e.
$\boldsymbol{S}\cdot\boldsymbol{S}=1$),
and $\boldsymbol{h}$ is an $N$-dimensional external vector field.
It is convenient to choose $\boldsymbol{h}=(h,0,\dots,0)$ by $O(N)$
symmetry and decompose $\boldsymbol{S}$ as $\boldsymbol{S}=(\sigma=\sqrt{1 -
  \pi^2}, \pi^i)$, where $\pi^i$ are the components of an $(N-1)$-dimensional vector
field with the condition $\pi^2 \leq 1$. 
Then the action can be rewritten as
\bea
\cS&=&\frac{1}{2g^2}\int d^2x
\left\{(\p_{\mu} \pi^i)^2
+\frac{(\pi \cdot \p_\mu \pi)^2}{1-\pi^2}
   -2h \sigma
\right\}
\nonumber \\
&=&
\frac{1}{2g^2}\int d^2x
\left\{(\p_{\mu} \pi^i)^2
+(\pi \cdot \p_\mu \pi)^2
+\cdots
   -2h + h \pi^2 +\frac{h}{4}(\pi^2)^2
+\cdots
\right\} \,,
\eea
where the external magnetic field $h$ plays the role of the infrared
regularization to give a mass to the $\pi$ fields.
We regard this as a massive counterpart of the original $O(N)$
nonlinear sigma model without the external field.

The 1PI two-point function $\Pi^{ij}$ of two $\pi$ fields at one-loop order is
\bea
\Pi^{ij}(q^2)
=
   -\delta^{ij}
\left[q^2+\frac{N-3}{4}h\right]h^{-\ep/2}(4\pi)^{\ep/2-1}
\Gamma(\ep/2) ,
\eea
where $d=2-\epsilon$.
It follows that the physical pole mass is
\bea
M^2 =h\left[
1-\frac{g^2}{4\pi}
h^{-\ep/2}\frac{N+1}{4}(4\pi)^{\ep/2}
\Gamma(\ep/2)
\right] .
\eea
It is easy to see that there is no non-logarithmic perturbative
correction at one-loop order.

\begin{figure}[t]
\begin{center}
\rotatebox{0}{\includegraphics[width=20em,clip]{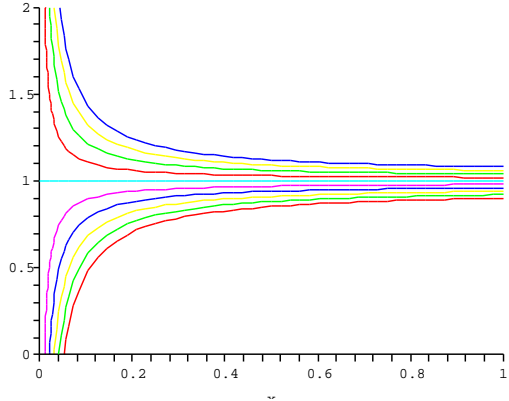}}
\end{center}
\caption{The plateau in the nonlinear sigma model.
The plateau emerges at $M_\text{P}^2/\Lambda_{\overline{MS}}^2=1$ when $a=1$.
The horizontal and the vertical axes denote $h$ and $M^2/\Lambda^2$,
respectively.}
\label{fig:NLS}
\end{figure}

We also obtain the renormalization functions
\bea
\beta(g^2)=-\frac{N-2}{2\pi}g^4 +O(g^6), \qquad
\gamma_m=\frac{N+1}{8\pi}g^2 +O(g^4) \,,
\eea
and thus we have
\bea
b_0=\frac{N-2}{4\pi}, \qquad
\gamma_0=\frac{N+1}{8\pi} .
\eea
Substituting these into the pole mass formula (\ref{polemassformula}),
we find the plateau (in the case $N=3$) depicted in
Fig.~\ref{fig:NLS}, in which
we have plotted $M_\text{P}^2/\Lambda_{\overline{MS}}^2$ with respect
to $h$ with various values of $a$.
We recognise that a plateau emerges when we set $a=1$, and the value on
the plateau is $M_\text{P}^2/\Lambda_{\overline{MS}}^2=1$.
This result is indeed identical to the exact value in the large-$N$ limit,
$M^2/\Lambda^2=1$.
Note that also in the Gross-Neveu model,
 the one-loop result is exact in the large-$N$ case \cite{AGN-AGIKN}.
However, it is rather far from the exact value for $N=3$,
 $M/\Lambda_{\overline{MS}}=8/e\simeq 2.94$.\footnote{
The expression for the exact mass gap evaluated in Ref. \citen{Hasenfratz:1990zz} is
\bea
M=\frac{(8/e)^{\frac{1}{N-2}}}{\Gamma(1+\frac{1}{N-2})}
\Lambda_{\overline{MS}} .
\eea
}
It can be seen that the deviation from the basic scale $\Lambda$ depends
on the value of the non-logarithmic perturbative correction.
It has been previously shown that the two-loop results for the Gross-Neveu
model exhibit good agreement with the exact results even
for smaller $N$ \cite{AGN-AGIKN}. 
Thus, we believe that if we proceed beyond one-loop order, we will
obtain a more accurate result in this model as well.

Let us comment on the spontaneous symmetry breaking here.
Usually a non-zero expectation value of a mean field after
the appropriate subtraction indicates that symmetry is broken
spontaneously.
However in the non-linear sigma model in two dimensions,
there never occurs a spontaneous symmetry breaking
because of its dimensionality.
Thus a non-zero value of a mean field does not
imply SSB.
Rather, it merely indicates that
the starting point of the perturbation theory does not respect the
symmetry.
In the non-linear
sigma model, the generated mass term is indeed $O(N)$ invariant,
 and thus SSB does not occur.
If the generated mass term breaks a symmetry,
then one can conclude that SSB takes place
and use the generated mass term as an order parameter,
as in the Gross-Neveu model, where a fermion mass term breaks the chiral
symmetry. 

\section{Improved RG analysis for Yang-Mills theory}
\label{sec:YM}

Let us start with a comment on what we are going to calculate using the
improved RG analysis in pure Yang-Mills theory.
The improved RG analysis presented here is basically a method to calculate a mass
term in a theory.
In Yang-Mills theory, the mass term for the gluon is known to be strictly
forbidden by gauge symmetry.
Indeed, it is shown in Ref. \citen{Barnich:ve} that there is no mass dimension two BRST
invariant local operator which can be a source of mass generation.
In a specific gauge, however, there can exist a local mass dimension two
operator which may be a source of mass generation, though this
``mass'' does not have any gauge-invariant meaning and is not a physical
mass.
Below we consider one of these operators, $\Tr A_\mu A^\mu$, in
the Landau gauge.
This is gauge dependent and becomes nonlocal in a general gauge.
However, as long as we start with the gauge-fixed action in the Landau
gauge, in which there is no classical gauge invariance, we can think
about the condensation of this operator, and through the four-point
interaction of gluons, a mass term for the gluon may be generated.
In the gauge-fixed action, BRST invariance is an important symmetry,
and therefore it has to
be maintained throughout calculation.
This operator, with a space-time integral, is indeed BRST
invariant in the Landau gauge.
In the following section, 
we consider the
Curci-Ferrari model as our massive gauge theory counterpart.
Although it seems at first sight that the mass term will undermine
the unitarity of the original Yang-Mills theory, our mass term 
is fictitious.
In the prescription of the improved RG method,
we introduce a counterterm which cancels the effect of the mass term.
Therefore, we expect that pathologies of the massive gauge theory will
 eventually disappear.

As we will see, the mass term has a nonperturbative nature; that is,
it is proportional to the scale given by the dimensional
transmutation.
A proportional coefficient would be of order 1.
This term might be
corrected by perturbative quantum effects, like the instanton action.
However, due to the lack of a clear understanding regarding the mass
generation and
condensation of this operator, it is difficult to calculate
corrections to determine the exact amount of the generated mass.
For this reason, we restrict our investigation to the leading order,
and the formulation of a precise 
argument for these coefficients is postponed to a future study.


The Lagrangian of the Curci-Ferrari model that is our
counterpart of $U(N)$ Yang-Mills theory is
\begin{align}
\cal{L}=& -\frac{1}{4}(\p_\mu A_\nu^a-\p_\nu A_\mu^a)^2 \nn\\
&
  -\frac{1}{2\xi}(\p_\mu A^{a\mu})^2
   -\frac{1}{2}g
f^{abc}A^\mu_aA^\nu_b(\p_\mu A_{\nu c}-\p_\nu A_{\mu c})
   -\frac{1}{4}g^2
f^{abc}f^{ade}A_{\mu b}A_{\nu c}A^{\mu}_dA^{\nu}_e \nn \\
& +\p_\mu\bar{c}^a\p^\mu c^a
+gf^{abc}(\p_\mu \bar{c}^a)A^{\mu b} c^c
-\frac{g}{2}(1-\beta)f^{abc}(\p^\mu A^a_\mu) \bar{c}^b c^c \nn \\
&+\frac{g^2}{8}\xi(1-\beta^2)f^{abc}f^{ade}
\bar{c}^b c^c \bar{c}^d c^e \nn\\
&+ \frac{1}{2}m^2 A_\mu^a A^{a\mu} -m_\text{gh}^2\bar{c}^ac^a \,,
\end{align}
where $\xi$ is the gauge parameter and $\beta$ is an extra parameter
characteristic of the model.
Here, the ghost mass is related to the gluon mass as $m_\text{gh}^2=\xi
m^2$.
The condition that the model possesses the BRST symmetry is given
by
\bea
\xi \beta m^2 =0 .
\eea
We carry out the calculation by setting $\beta=1$.
The BRST invariance will be manifest in the last stage of the calculation,
where we could regard the original mass as going to zero as a result of the
subtraction or when we go to the Landau gauge, $\xi=0$.

To find the perturbative expression for the pole mass of the theory,
it is sufficient to calculate the one-particle irreducible (1PI) two-point
function.
We write the 1PI two-point function as follows:
\bea
i\Pi_{ab}^{\mu\nu}
\equiv
i\delta_{ab}
\left[
\pi_1 m^2 \eta^{\mu\nu}+\pi_2 q^2 \Delta^{\mu\nu}
\right] ,
\eea
where $\Delta^{\mu\nu} = \eta^{\mu\nu}-q^\mu q^\nu/q^2$,
and $\pi_1$ and $\pi_2$ should be calculated perturbatively.
Then, the full propagator $i\Sigma^{\mu\nu}$ is written
in the form
\bea
i\Sigma^{\mu\nu} =
\frac{-i/(1-\pi_1)}{q^2-\frac{1+\pi_1}{1-\pi_2}m^2}
\left(
\eta^{\mu\nu}-\frac{q^\mu q^\nu}{\frac{1+\pi_1}{1-\pi_2}m^2}
\right)
+\frac{-i/(1-\pi_1)}{q^2-(1+\pi_1)m^2}\cdot
\frac{q^\mu q^\nu}{\frac{1+\pi_1}{1-\pi_2}m^2} .
\eea
Thus, the pole mass can be read off as
\bea
M_P^2=m^2 \frac{1+\pi_1}{1-\pi_2}
= m^2 (1+\pi_1+\pi_2) +O(g^4) .
\eea

The 1PI two-point function is calculated up to one-loop order
in a standard way as
\bea
i\Pi^{ab}_{\mu\nu} &=&
\frac{ig^2 N \delta^{ab}}{(4\pi)^{2}}
\int_0^1 dx\int_0^1 dy\int_0^1 dz
\left[
\left(\frac{2}{\bar{\ep}}
\right)(m^2\eta^{\mu\nu}N_{m0}+q^2\Delta^{\mu\nu}N_{A0})
\right.
\nonumber \\
&&
\left.
   -\ln\Delta
(m^2\eta^{\mu\nu}N_{m0}+q^2\Delta^{\mu\nu}N_{A0})
+2(m^2\eta^{\mu\nu}N_{m1}+q^2\Delta^{\mu\nu}N_{A1})
\right] ,
\label{1PI}
\eea
where $\bar{\epsilon}^{-1} = \epsilon^{-1} - \gamma +\ln (4\pi)$, 
\bea
\Delta =\left[1-(1-\xi)\left\{zx+(1-z)y\right\}\right]m^2-z(1-z)q^2 ,
\eea
and the explicit expressions for the coefficient functions
$N_{m0}(x,y,z;\xi)$, $N_{m1}(x,y,z,m^2,q^2;\xi)$, 
$N_{A0}(x,y,z;\xi)$ and $N_{A1}(x,y,z,m^2,q^2;\xi)$
are listed in Appendix \ref{Ns}.

The constant $A_{1}$ is given by the finite part of
$\pi_1+\pi_2$, which is found from (\ref{1PI}) to be
\bea
A_{1}=
\frac{g^2 N}{(4\pi)^2}
\left[
\frac{313}{36}-\frac{11\sqrt{3}\pi}{8}-\frac{7}{12}\xi
+\frac{(4\xi-1)^{3/2}}{12}\arctan \left(\frac{1}{\sqrt{4\xi-1}}\right)
+\frac{6\xi-1}{24}\ln \xi
\right] .
\nonumber \\
\eea
In the Landau gauge, we have $\xi \to 0$, and $A_1$ becomes
\bea
\label{A_1}
A_1=\frac{g^2N}{(4\pi)^2}\left[\frac{313}{36}-\frac{11\sqrt{3}\pi}{8}\right] .
\eea
The renormalization functions are
\bea
\beta(g^2)=-\frac{11}{3}N\frac{g^4}{(4\pi)^2} +O(g^6), \qquad
\gamma_m=\frac{35-3\xi}{12}N\frac{g^2}{(4\pi)^2} +O(g^4) \,,
\eea
and thus
\bea
\label{renorm_const_YM}
b_0=\frac{11}{6}\frac{N}{(4\pi)^2} , \qquad
\gamma_0=\frac{35-3\xi}{12}\frac{N}{(4\pi)^2}
\eea

\begin{figure}[t]
\begin{center}
\rotatebox{0}{\includegraphics[width=20em,clip]{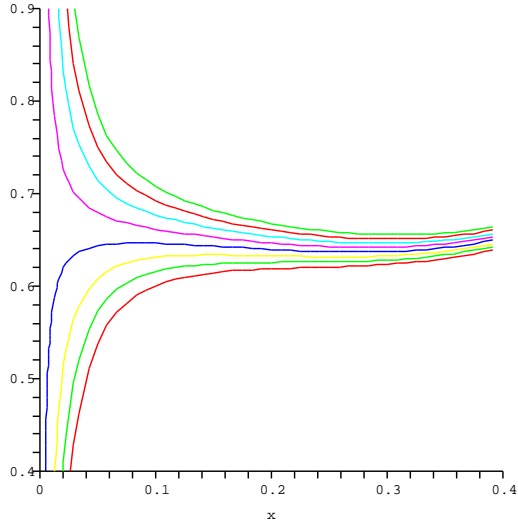}}
\end{center}
\caption{The plateau in the Landau  gauge.
The vertical axis represents $M_\text{P}^2/\Lambda_{\overline{MS}}^2$,
evaluated using (\ref{eq:IRG_formula}) with the data (\ref{A_1}) and
(\ref{renorm_const_YM}).
The horizontal axis represents $x$ proportional to $m^2$.
We have plotted the graphs for various values of $a$ from $a=0.6565$ to $a=0.6635$.
The plateau appears near
$M_\text{P}^2/\Lambda_{\overline{MS}}^2\simeq 0.66$.}
\label{fig:Landaugauge}
\end{figure}

Using the above results, we apply the improved renormalization group
method explained above to Yang-Mills theory and are able to
calculate the approximated pole mass of the gluon.
Note that at one-loop order, the result is independent of $N$.
We plot $M_\text{P}^2/\Lambda^2_{\overline{MS}}$ with respect to $x$,
which corresponds to the provisional mass $m^2$, for various
values of $a$ in Figure \ref{fig:Landaugauge}.
Figure \ref{fig:Landaugauge} shows that a plateau emerges near
$M_\text{P}^2/\Lambda^2_{\overline{MS}} \simeq 0.66$.

%


\section{Conclusion}
\label{sec:conc}

In this paper, we have presented one method for evaluating the operator 
$\Delta=\frac{1}{2}(\text{volume})^{-1}\langle \text{min}_U \int d^4 x
(A_\mu^U)^2 \rangle$.
In the Landau gauge, the calculation boils down to evaluating $\langle A_\mu
A^\mu \rangle$, which may be connected to the mass pole of the gluon
propagator in this gauge.
We have thus applied the improved renormalization group method to pure
Yang-Mills theory at one-loop order and obtained the result that a
nonperturbatively
generated pole mass of the gluon, $M_\text{P}$, emerges as
$M_\text{P}^2/\Lambda_{\overline{MS}}^2 \simeq 0.66$, where
$\Lambda_{\overline{MS}}$ is the $\overline{MS}$ scale.
Some work has been done to calculate the mass of the gluon, for
example, in a lattice calculation \cite{Boucaud:2001st}.
Our result is much smaller than those found in the previous work.
However, we cannot compare the previous result with our result directly, because that
result is not the pole mass itself but, rather, the vacuum expectation value
of the $A_\mu A^\mu$ operator calculated through the operator product expansion.
It would be interesting to perform further calculations, two loops or higher,
and observe if higher-order calculations might give larger gluon
masses.
Calculations of the renormalization functions
of the Curci-Ferrari model have been carried out to three-loop order
\cite{G-G-BG}.
Those calculations enable
us to improve the renormalization group analysis, but we think that
perturbative corrections play an important role in producing a
nonperturbative mass.
Again, we would like to emphasize that though we have done the
calculation in the Landau gauge, the result should have a physical meaning,
because the original $\Delta$ is a gauge-invariant quantity.

Here, let us comment on a possible relationship between mass and confinement.
If the pole mass of the gluon were infinitely large, we could interpret
it as a signal of confinement, since such a large mass would indicate that
the gluon does not propagate as a physical mode.
This is an interesting viewpoint,\footnote{The authors owe discussion
  on this viewpoint to Prof.~Kawai.} and we expected to find a
large gluon mass.
Although one-loop order we have not obtained such a large mass, the
gluon mass might become larger and larger as the order of
calculation increases.

Yet another perspective concerning confinement may exist.
In the operator formalism of Yang-Mills theory, there is a well-known
criterion for confinement, the so-called Kugo-Ojima
condition\cite{Kugo:1979gm}, which
is a sufficient condition for color confinement.
The condition mentioned here concerns a BRST transformation property of
anti-ghost fields, and the criterion
asserts that when this condition is satisfied, one can prove that 
no color singlet state belongs to the
BRST cohomology, and hence it cannot be observed as a physical state.
This condition has been interpreted as involving the
infrared properties of gluon and
ghost propagators in the Landau gauge\cite{Kugo:1995km},
and these properties have been tested with lattice
simulations\cite{Furui-Nakajima},
analytically\cite{Watson:2001yv} and by employing the Schwinger-Dyson
equations\cite{Alkofer:2001iw}.
Because the infrared behavior of the propagator is involved with the
existence of the nonperturbatively generated mass, it would be interesting
to study it with our method.\footnote{%
The authors thank an anonymous referee for bringing this interesting
idea to their attention.
For a review of the infrared properties of QCD
propagators,
see Ref.~\citen{Alkofer:2000wg}.}
This viewpoint also provides an interesting interpretation of the remark
above, that is, that the confinement via the Kugo-Ojima mechanism
requires a finite mass for gluon propagators, and thus it may be
consistent with our result.

Another issue concerns the ghost mass generation.
In the literature, generation of the ghost mass has been studied
intensively,
and it might also be possible to evaluate it using
 the prescription presented in this paper.
It would be interesting to calculate the pole mass of the ghost propagator.

We have also applied this method to the $O(N)$ non-linear sigma model in two dimensions.
The exact mass gap of this model is known
\cite{Hasenfratz:1990zz}, and we can compare it with our result in
order to check the effectiveness of this method.
However, at one-loop order, the calculation is consistent with the exact
result only in the large-$N$ limit, where we have $M_\text{P} = \Lambda_{\overline{MS}}$.
A similar fact  was observed in
the Gross-Neveu model \cite{AGN-AGIKN} at one-loop order.
In Ref.~\citen{AGN-AGIKN}, a two-loop
calculation is performed and a result in good agreement with the exact result,
even for small values of $N$, is obtained.
Considering this fact, it would be interesting to calculate the next
order in the non-linear sigma model to make sure of the validity of this
method.


\section*{Acknowledgements}

The authors would like to thank H.~Kawai for helpful discussions
and valuable comments.
They also thank T.~Kuroki and the members of the theoretical physics
laboratory of RIKEN for discussions and useful comments.
T.M would like to thank K. Kondo for comments and
K. Higashijima for his pointing out an error in a calculation in a
preliminary draft.
T.M is especially grateful to Y. Shibusa for discussions.
The authors thank Prof. J. A. Gracey for his comments and for answering
their questions.
This work is supported by Special Postdoctoral Researchers Program at
RIKEN.

\appendix
\section{Coefficient Functions}
\label{Ns}
Below we give explicit expressions for the coefficient functions
$N_{m0}, N_{A0}, N_{m1}$ and $N_{A1}$:
\begin{align}
N_{m0} =&\frac{3}{2}z(1-z)
\left(16(1-\xi)zx+16(1-\xi)(1-z)y-5+9\xi\right) \,,
\\
N_{A0} =&z(1-z)(80z(1-z)-3(1+\xi)) \,,
\\
N_{m1} =&
\left[
\left\{
   -19(1-\xi)^3 z(1-z)((1-z)y+zx)^3
\right.\right.\nn\\&\left.\left.\hskip0.18cm
   -\frac{1}{4}(63\xi-181)(1-\xi)^2z(1-z)((1-z)y+zx)^2
\right.\right.\nn \\ & \left.\hskip0.2cm
   -\frac{3}{4}(1-\xi)(2\xi^2-35\xi+45)z(1-z)((1-z)y+zx)
\right.\nn \\ & \left.\hskip0.2cm
   -\frac{2}{3}(1-\xi)(2\xi-5)z(1-z)
\right\}(m^2)^2
\nn \\ &\hskip0.24cm  +
\left\{
(1-\xi)^2(-46z(1-z)+3)z(1-z)((1-z)y+zx)^2
\right. \nn \\ &\left.\hskip0.22cm
+\frac{1}{8}(1-\xi)
\right.\nn\\&\left.\hskip0.8cm \times
(-174\xi z(1-z)+446z(1-z)-3\xi-21)z(1-z)((1-z)y+zx)
\right. \nn \\ &\left.\hskip0.22cm
+\frac{1}{8}z(1-z)(-24\xi^2 z(1-z)-76z(1-z)+156\xi z(1-z)-9\xi-3)
\right\}m^2 q^2 \nn \\ &\hskip0.24cm +
\left\{
(1-\xi)(-26z(1-z)+3)z^2(1-z)^2((1-z)y+zx)
\right. \nn \\ &\left.\left.\hskip0.21cm 
+\frac{1}{8}z^2(1-z)^2(-72\xi z(1-z)+32z(1-z)-1+9\xi)
\right\}(q^2)^2
\right]/\Delta^2 \,,
\\
N_{A1} =&
\left[
\left\{
\frac{1}{2}(-116z(1-z)+7\xi+10)z(1-z)((1-z)y+zx)^2
\right.\right. \nn \\& \left.\hskip0.5cm
+(1-\xi)(-22\xi z(1-z)+122z(1-z)+\xi^2-7\xi-11)
\right.\nn\\&\left.\hskip1cm
\times z(1-z)((1-z)y+zx)
\right.
\nn \\&\left.\hskip0.5cm
+\frac{1}{4}z(1-z)(12\xi^2z(1-z)-258z(1-z)
\right.\nn\\&\left.\hskip0.15cm\phantom{\frac{1}{2}}
+86\xi z(1-z)-3\xi^2+15\xi
+24)
\right\}(m^2)^2
\nn \\ &-
\left\{
\frac{1}{8}(1-\xi)^2z(1-z)
(4z(1-z)(272z(1-z)-33)
\right. \nn \\&\left.\hskip0.5cm
   -28\xi z(1-z)-\xi^2-4\xi-3)((1-z)y+zx)
\right. \nn \\ &\left.\hskip0.36cm
+\frac{1}{8}z(1-z)(4z(1-z)(281z(1-z)-34)+4\xi z(1-z)(33z(1-z)+10)
\right. \nn \\ &\left.\phantom{\frac{1}{2}}
+8\xi^2 z(1-z)+3\xi^2+6\xi+3)
\right\}m^2 q^2
\nn \\ &\hskip0.4cm +\left.
\frac{1}{8}z^2(1-z)^2(8\xi z(1-z)+\xi^2+4\xi
\right.\nn\\&\left.\hskip0.38cm
+3-8z(1-z)(76z(1-z)-11))
(q^2)^2
\right]/\Delta^2 .
\end{align}


\section{Renormalization Group Analysis for the Pole Mass}
\label{LeadingLog}

In this appendix, we explain the fact that a pole mass can be
written in terms of a logarithm of $m^2/\mu^2$, following Ref.~\citen{Collins}.
We also comment on a leading-logarithm approximation.

Here we concentrate on a theory that has two dimensional parameters,
$m^2$ and $\mu$.
The physical mass that appears at a pole of the propagator is
\begin{equation}
\label{Mpexp1}
      M_{\text{P}}^2
      =m^2 \sum_{n=0}^{\infty} f_n \left(\frac{m^2}{\mu^2}
      \right) g^{2n} ,
\end{equation}
where the quantities $f_n$ are undetermined functions of $m^2/\mu^2$ and are
calculated by perturbation theory, and $g$ is the coupling constant of
the theory.
Note that here we have applied a mass-independent renormalization (MIR)
scheme, like minimal subtraction $MS$ or $\overline{MS}$.
Only when an MIR scheme is applied does the pole mass have a form
like (\ref{Mpexp1}).

By definition, the pole mass $M_{\text{P}}^2$
is independent of $\mu$; that is, it is renormalization group
invariant.
Then, $M_{\text{P}}^2$ satisfies the renormalization group equation
(RGE)
\begin{equation}
      \left[ \mu \dder{\mu} + \beta(g)\dder{g^2}
      - \gamma_{m^2}(g) m^2
      \dder{m^2} \right] M_{\text{P}}^2=0 ,
\end{equation}
where
\begin{equation}
      \beta(g)= \mu \drdr{g^2}{\mu}, \ \
      \gamma_{m^2}=-\frac{\mu}{m^2} \drdr{m^2}{\mu} .
\end{equation}
Substituting (\ref{Mpexp1}) into this RGE, we obtain
\begin{equation}
      \dder{\ln \mu} f_n = \left[ \sum_{n'<n}
        \left(  \gamma_{m^2} m^2 \dder{m^2} - \beta
        \dder{g^2} - \gamma_{m^2} \right) f_{n'} g^{2n'}
        \right]_{\text{coeff. of $g^{2n}$}} .
\end{equation}
Integrating with respect to $\ln \mu$, we obtain
\bea
      f_n&=&\left[ \sum_{n'<n}
        \left(  \gamma_{m^2}m^2 \dder{m^2} - \beta
        \dder{g^2} - \gamma_{m^2} \right)
        \int_0^{\ln \mu} d\ln \mu' f_{n'}\left(\frac{m^2}{\mu^{'2}}
        \right) g^{2n'}
        \right]_{\text{coeff. of $g^{2n}$}}
        \nonumber \\
       && + \text{const.} \,,
\eea
where const. is independent of $\mu$ and, thus, $m^2$.
We have the ``boundary condition'' $f_0=1$ of this integral equation, 
and hence
\begin{equation}
      f_1= \text{const.} -\frac{1}{2} \gamma_0 \ln \left(
        \frac{m^2}{\mu^2} \right) ,
\end{equation}
where we have defined
\begin{equation}
      \beta(g)=-2b_0 g^4 -2b_1 g^6 + \cdots, \ \
      \gamma_{m^2}=
      \gamma_0 g^2+ \gamma_1 g^4+ \cdots .
\end{equation}
In this manner, we can determine all the functions $f_n$ recursively by using
of information provided by the RGE, up to constant terms.
Thus we obtain
\begin{equation}
\label{Mplog}
      M_\text{P}^2 = m^2 \sum_{n=0}^{\infty} \sum_{L=0}^n A_{n,L}
      \left[\ln \left(\frac{m^2}{\mu^2} \right)\right]^{n-L} g^{2n},
\end{equation}
where the coefficients $A_{n,L}$ with $L<n$ are determined through the
above procedure, while $A_{n,n}$($\equiv A_n$) are unknown constants at
this stage.
These constants are to be calculated within the perturbation theory,
and we call them ``non-logarithmic corrections''.

An interesting fact is that the $L=0$ contribution in the series
(\ref{Mplog}), called a leading logarithm approximation,
satisfies the renormalization group equation at one-loop order,
that is,
\begin{gather}
       \left[ \mu \dder{\mu} -2b_0 g^2\dder{g^2}
       +  \gamma_0 g^2 m^2\dder{m^2} \right] 
     \left( M_{\text{P}}^\text{LL} \right)^2
       =0 .
\end{gather}
Therefore we can solve it in a standard way, and we obtain
\begin{equation}
     \left( M_\text{P}^\text{LL} \right)^2 = m^2
     \sum_{n=0}^{\infty} A_{n,0} \left[ \ln \left(
         \frac{m^2}{\mu^2} \right) \right]^n g^{2n} .
\end{equation}
This indicates that once we know $b_0$ and $\gamma_0$ at
one-loop order, we can calculate the leading-logarithm contribution to
the pole mass at all orders in $g^2$.


\end{document}